\documentclass[11pt]{article}

\usepackage[utf8]{inputenc}
\usepackage{subfigure}
\usepackage{graphicx}
\usepackage{amsmath}
\usepackage{authblk}
\usepackage{fullpage}
\usepackage[sort, round]{natbib}

\begin{document}

\title{Intermittency and transition to chaos in the cubical lid-driven cavity flow}

\author[1]{J-Ch. Loiseau}
\author[2]{J-Ch. Robinet}
\author[3]{E. Leriche}

\affil[1]{Department of Mechanics, Royal Institute of Technology (KTH), SE-100 44 Stockholm, SWEDEN}
\affil[2]{Laboratoire DynFluid, Arts et M\'etiers ParisTech, 75013 Paris, FRANCE}
\affil[3]{Laboratoire de M\'ecanique de Lille, Universit\'e Lille 1, 59655 Villeneuve d'Ascq, FRANCE}

\maketitle

\begin{abstract}

Transition from steady state to intermittent chaos in the cubical lid-driven flow is investigated numerically. Fully three-dimensional stability analyses have revealed that the flow experiences an Andronov-Poincar\'e-Hopf bifurcation at a critical Reynolds number $Re_c = 1914$. As for the 2D-periodic lid-driven cavity flows, the unstable mode originates from a centrifugal instability of the primary vortex core. A Reynolds-Orr analysis reveals that the unstable perturbation relies on a combination of the lift-up and anti lift-up mechanisms to extract its energy from the base flow. Once linearly unstable, direct numerical simulations show that the flow is driven toward a primary limit cycle before eventually exhibiting intermittent chaotic dynamics. Though only one eigenpair of the linearized Navier-Stokes operator is unstable, the dynamics during the intermittencies are surprisingly well characterized by one of the stable eigenpairs.

\end{abstract}

\vspace{2pc}
\noindent{\it Keywords}: Lid-driven cavity, global instability, intermittency, chaos.

\section{Introduction}

The two- and three-dimensional lid-driven cavity flows are canonical problems of fluid dynamics qualitatively presenting some of the key features responsible for transition to turbulence in a wide variety of confined flow situations. Over the past twenty years, numerous studies have investigated the linear stability properties of the two-dimensional lid-driven cavity flow~\citep{Ramanan_POF_1994,Ding_IJNMF_1998,Albensoeder_PoF_2001,Theofilis_JFM_2004,Non_POF_2006,Jerem_CRAS}. Depending on the Reynolds number and the spanwise wavenumber of the prescribed perturbation, the 2D-periodic LDC flow is unstable toward four different families of modes~\citep{Theofilis_JFM_2004}. The first bifurcation for the square lid-driven cavity occurs at a critical Reynolds number $Re_{S1} = 780$ for a spanwise wavenumber $\beta \simeq 15.4$. The associated branch is known as the S1 family of modes. It is a family of non-oscillating Taylor-G\"ortler-like (TGL) vortices. Further increasing the Reynolds number drives a larger range of wavenumbers to become unstable and the flow eventually experiences an Andronov-Poincar\'e-Hopf bifurcation (T1 family) yielding it to transition to unsteadiness beyond a critical Reynolds number $Re_{T1} = 840$. Qualitatively similar results have been obtained regarding the 2D-periodic stability of the shear-driven open cavity flow~\citep{Meseguer_JFM_2014,Citro_JFM_2015}. All these families of modes are related to a centrifugal instability of the primary vortex core~\citep{Albensoeder_PoF_2001}.

\bigskip

The flow within a three-dimensional enclosure has received much less attention than its two-dimensional counterpart. In the mid 1970's, \cite{Davis_CF_1976} started to study this three-dimensional setup. The flow developing within three-dimensional cavities qualitatively exhibits the same features as its two-dimensional counterpart: a central primary vortex flanked with edge eddies. For extensive details about the topology of three-dimensional LDC flows, the reader is referred to the in-depth review by \cite{Shankar_Deshpande_LDC}. \cite{Albensoeder_JCP_2005} have provided in 2005 accurate data of the steady flow within a cubical LDC at low Reynolds number ($Re=1000$) using a Chebyschev collocation method while, in 2007, \cite{Bouffanais_PoF_2007} have investigated the dynamics of the flow at high Reynolds number ($Re=12000$) using Legendre spectral elements (see also \cite{Leriche_PoF_2000}). However, because investigating the linear stability analysis of fully three-dimensional flows still is computationally challenging, relatively few references can be found on the linear instability and transition thresholds. In 2010, \cite{Feldman_PoF_2010} addressed the stability of the cubical lid-driven cavity by means of direct numerical simulations. The flow transitions to unsteadiness at a critical Reynolds number $Re_c = 1914$ via an Andronov-Poincar\'e-Hopf bifurcation. As for the 2D-periodic stability analysis, the exponentially growing perturbation in their DNS takes the form of Taylor-G\"ortler-like vortices. In 2011, \cite{Liberzon_PoF_2011} conducted an experimental investigation on the cubical setup. Beside a slight disagreement on the value of the critical Reynolds number, both the frequency and the {\it rms} fluctuations in the experiment are in good agreements with the numerical predictions by \cite{Feldman_PoF_2010}. Quite recently, \cite{Kuhlmann_PoF_2014} and \cite{Gomez_AST_2014} have also investigated this problem and found similar critical Reynolds numbers. Moreover, once the flow is linearly unstable, \cite{Kuhlmann_PoF_2014} have observed that it is driven toward a primary limit cycle eventually exhibiting intermittent dynamics. Such intermittencies are typical of a dynamical system exhibiting chaotic dynamics and have been observed in a large variety of problems~\citep{Broze_JFM_1996,Kabiraj_JFM_2012}. Based on their observations, \cite{Kuhlmann_PoF_2014} suggest that the chaotic intermittent dynamics observed in the cubical lid-driven cavity flow could be related to the type-II intermittency in the classification by \cite{Pomeau_Manneville}. Due to the very long time integration necessary to fully characterize these dynamics, their investigations have however not been pushed further.

\bigskip

Though a complete characterization of these intermittencies has not been possible yet, the present work aims at shedding some more light on these dynamics. The paper is structured as follows: first, the problem under consideration is presented in section~\ref{sec: Problem statement} along with a brief overview of the numerical methods used. Section~\ref{sec: Results} summarizes the results of the present investigation, while conclusive remarks and possible perspectives to this work are given in section~\ref{sec: Conclusion}.

\section{Problem statement}
\label{sec: Problem statement}

\subsection{Governing equations}

The motion of a newtonian fluid contained within a cubical enclosure of characteristic length $L$ and driven by a moving lid is considered. The flow is governed by the three-dimensional incompressible Navier-Stokes equations
~
\begin{eqnarray}
    \displaystyle \frac{\partial {\bf U}}{\partial t} + ({\bf U} \cdot \nabla) {\bf U} = -\nabla P + \frac{1}{Re} \Delta {\bf U} \\
    \nabla \cdot {\bf U} = 0
\label{eq: Navier-Stokes}
\end{eqnarray}

\noindent where ${\bf U} = (U,V,W)^T$ is the velocity vector and $P$ the pressure. Dimensionless variables are defined with respect to the characteristic length $L$ of the cavity and the constant velocity $U_0$ of the moving lid. Therefore, the Reynolds number is defined as $Re = U_0L/\nu$, with $\nu$ being the kinematic viscosity. The origin of the system of axes is assigned to be the geometrical center of the cavity such that the non-dimensional domain considered is $\mathcal{V} = \left[ -0.5, 0.5 \right]^3$. Except on the lid (for which $U(x,y=0.5,z) = U_0$), no-slip boundary conditions are applied on all the walls of the cavity. Finally, contrary to \cite{Botella_CF_1998} on the 2D lid-driven cavity or to \cite{Bouffanais_PoF_2007}, \cite{Albensoeder_JCP_2005} and \cite{Kuhlmann_PoF_2014} on the cubical one, no specific regularization or decomposition of the lowest-order singularities have been used at the edges between the lid and the non-moving walls. Nonetheless, extremely good agreement is obtained with the reference data of \cite{Albensoeder_JCP_2005} on the cubical LDC flow at $Re=1000$ \citep{Loiseau_PhD_2014}.

\subsection{Linear stability analysis}
\label{subsec: 3D stability}

The asymptotic dynamics of infinitesimal perturbations evolving in the vicinity of a given fixed point ${\bf U}_b$ can be predicted by linear stability analysis. In the present work, these dynamics are governed by the three-dimensional linearized Navier-Stokes equations
~
\begin{eqnarray}
  \displaystyle \frac{\partial {\bf u}}{\partial t} =  -({\bf{u}} \cdot \nabla) {\bf{U}}_b - ({\bf{U}}_b \cdot \nabla) {\bf{u}}-\nabla p + \frac{1}{
Re} \Delta {\bf{u}} \\
  \nabla \cdot {\bf{u}} = 0
\end{eqnarray}

\noindent where ${\bf u}=(u,v,w)^T$ is the perturbation velocity vector and $p$ the perturbation pressure. The boundary conditions are the same as system~\eqref{eq: Navier-Stokes} except on the moving lid where a zero-velocity condition is now prescribed. Introducing the perturbation state vector ${\bf q} = ({\bf u},p)^T$, this set of equations can be recast into the following time-autonomous linear dynamical system
~
\begin{equation}
  {\bf{B}} \frac{\partial {\bf{q}}}{\partial t} = {\bf{Jq}}
  \label{eq: dynamical system}
\end{equation}

\noindent where ${\bf{B}}$ is a singular mass matrix and ${\bf{J}}$ is the Jacobian matrix of the Navier-Stokes equations~\eqref{eq: Navier-Stokes} linearized around the fixed point ${\bf U}_b$. Unfortunatey, because of the extremely large number of degrees of freedom involved in the computation, solving the generalized eigenvalue problem associated to this fully three-dimensional linear dynamical system using standard algorithms is hardly possible at the present time. As a consequence, a {\it time-stepping approach}, popularized by \cite{Edwards_JCP_1994} and \cite{Bagheri_AIAA_2009}, is used. This approach relies on the fact that, once projected onto a divergence-free vector space, system~\eqref{eq: dynamical system} reduces to

\begin{equation}
  \frac{\partial {\bf{u}}}{\partial t} = {\bf{Au}}
  \label{eq: projected dynamical system}
\end{equation}

\noindent with ${\bf{A}}$ being the projection of the Jacobian matrix ${\bf{J}}$ onto the divergence-free vector space. Introducing the exponential propagator ${\bf{M}} = e^{{\bf A}\Delta T}$, one obtains the following eigenvalue problem

\begin{equation}
  \mu \hat{\bf{u}} = {\bf{M}}\hat{\bf{u}}
  \label{eq: EVP 2}
\end{equation}

\noindent The linear stability of the fixed point ${\bf U}_b$ is then governed by the eigenvalue $\mu$: if $\| \mu \| < 1$ then ${\bf U}_b$ is linearly stable, otherwise, if $\| \mu \| > 1$, it is linearly unstable. Though the exponential propagator ${\bf M}$ cannot be explicitly computed, its action onto a given vector can be easily approximated by time-marching the linearized Navier-Stokes equations from $t=0$ to $t=\Delta T$. This property hence allows us to use Arnoldi-based iterative eigenvalue solvers. Finally, the eigenpairs $(\mu,\hat{\bf u})$ of the exponential propagator ${\bf M}$ are related to those of the Jacobian matrix ${\bf J}$ by

\begin{eqnarray}
  \lambda = \displaystyle \frac{\log(\mu)}{\Delta T} \\
  {\bf B}\hat{\bf q} = \hat{\bf u}
\end{eqnarray}

\bigskip

Calculations have been performed using the code \textsc{Nek5000}~\citep{nek5000_site}. Spatial discretisation is done by a Legendre spectral elements method with polynomials of order $6$. The number of spectral elements is set to $10$ in each direction resulting in a total of $1000$ spectral elements for the whole domain. The convective terms are advanced in time using an extrapolation of order $3$, whereas a backward differentiation of order $3$ is used for the viscous terms resulting in the time-advancement scheme known as BFD3/EXT3. For more details about the spectral elements method, the reader is refered to \cite{Deville_SEM} and \cite{Karniadakis_SEM}. The numerical implementation of the Krylov-Schur decomposition used to solve the time-stepper formulation of the eigenvalue problem in the present work relies on the basic Arnoldi factorization presented in \cite{Loiseau_PhD_2014} and \cite{Loiseau_JFM_2014} and on the LAPACK library~\citep{LAPACK} for the linear algebra computations (Schur and Eigenvalue decompositions). It has been cross-validated with the initial Arnoldi implementation on a number of standard benchmarks available from the literature ({\it e.g.} 2D LDC flow, 2D cylinder flow, cubical LDC flow, etc). Linear stability results have been obtained using a Krylov subspace of dimension $m=96$ and a sampling period $\Delta T = 1$ enabling good convergence of the eigenvalues with circular frequency $\omega \le 1.5$.

\section{Results}
\label{sec: Results}

\subsection{Linear stability analysis} \label{subsec: linear stability analysis}

\begin{figure}
  \centering
  \subfigure[]{\includegraphics[width=.75\columnwidth]{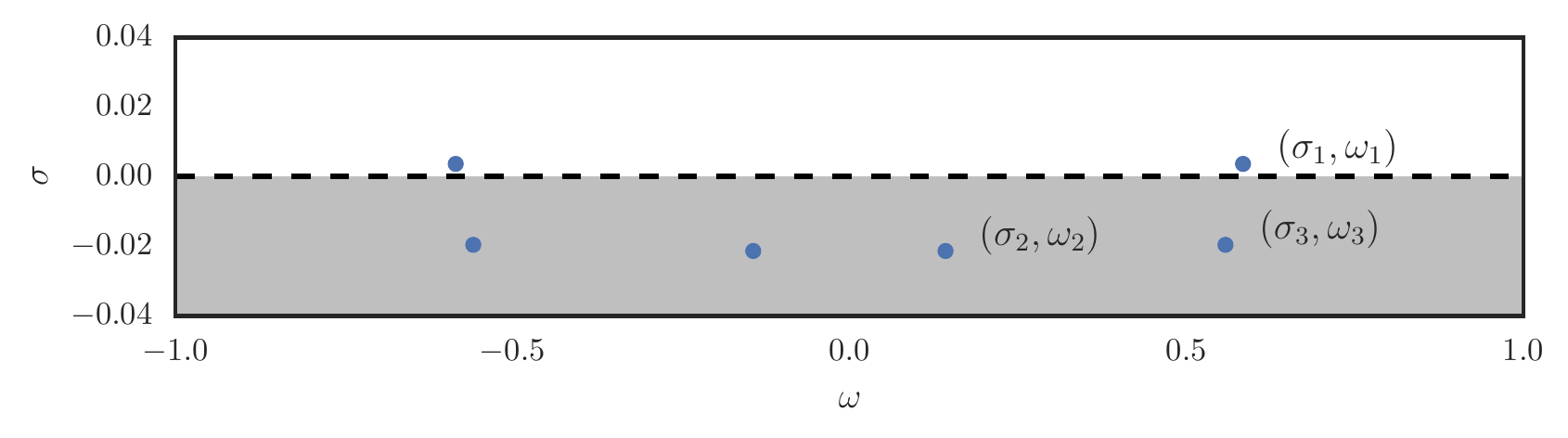}}
  \subfigure[]{\includegraphics[width=.4\columnwidth]{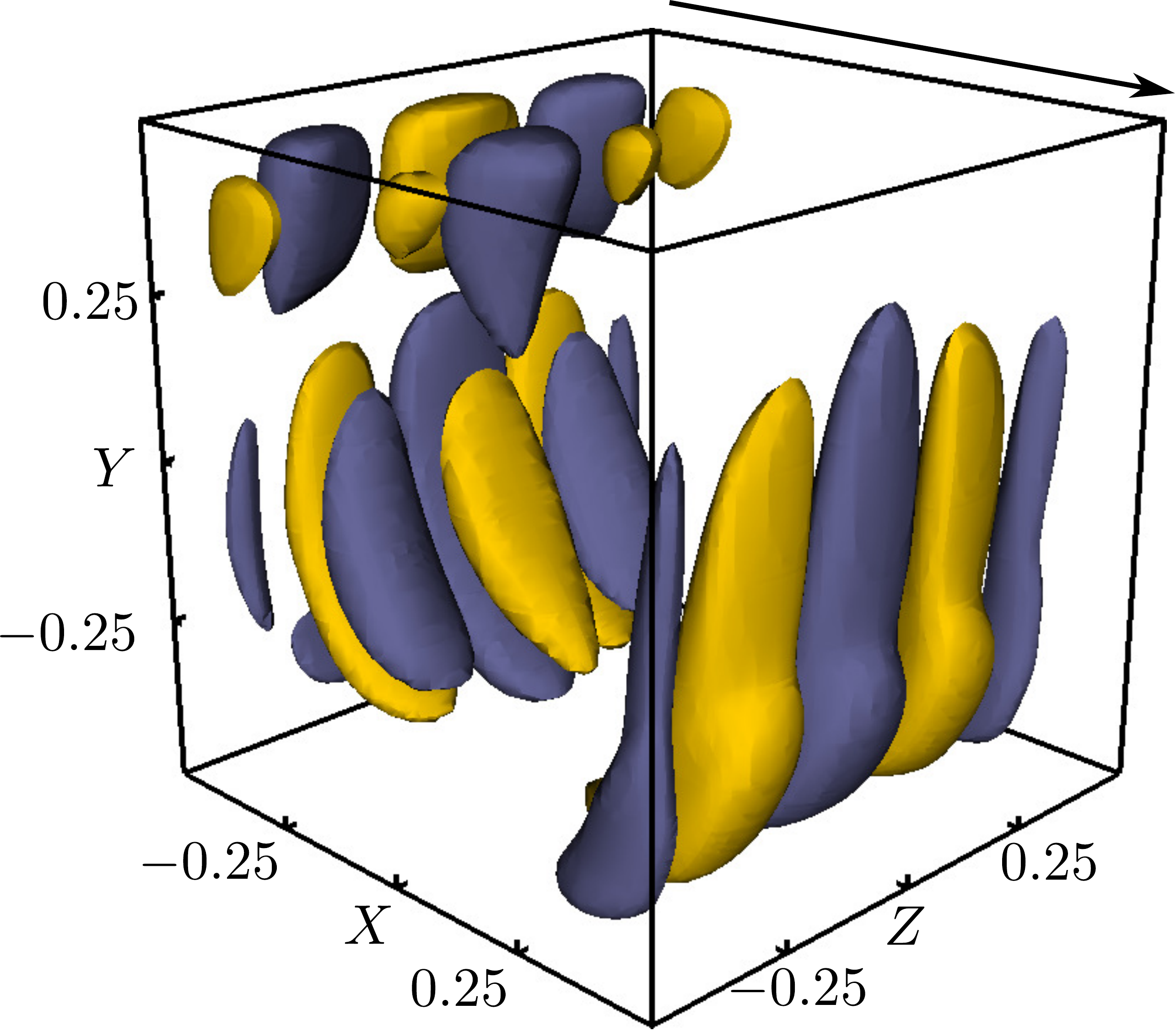}}
  \subfigure[]{\includegraphics[width=.35\columnwidth]{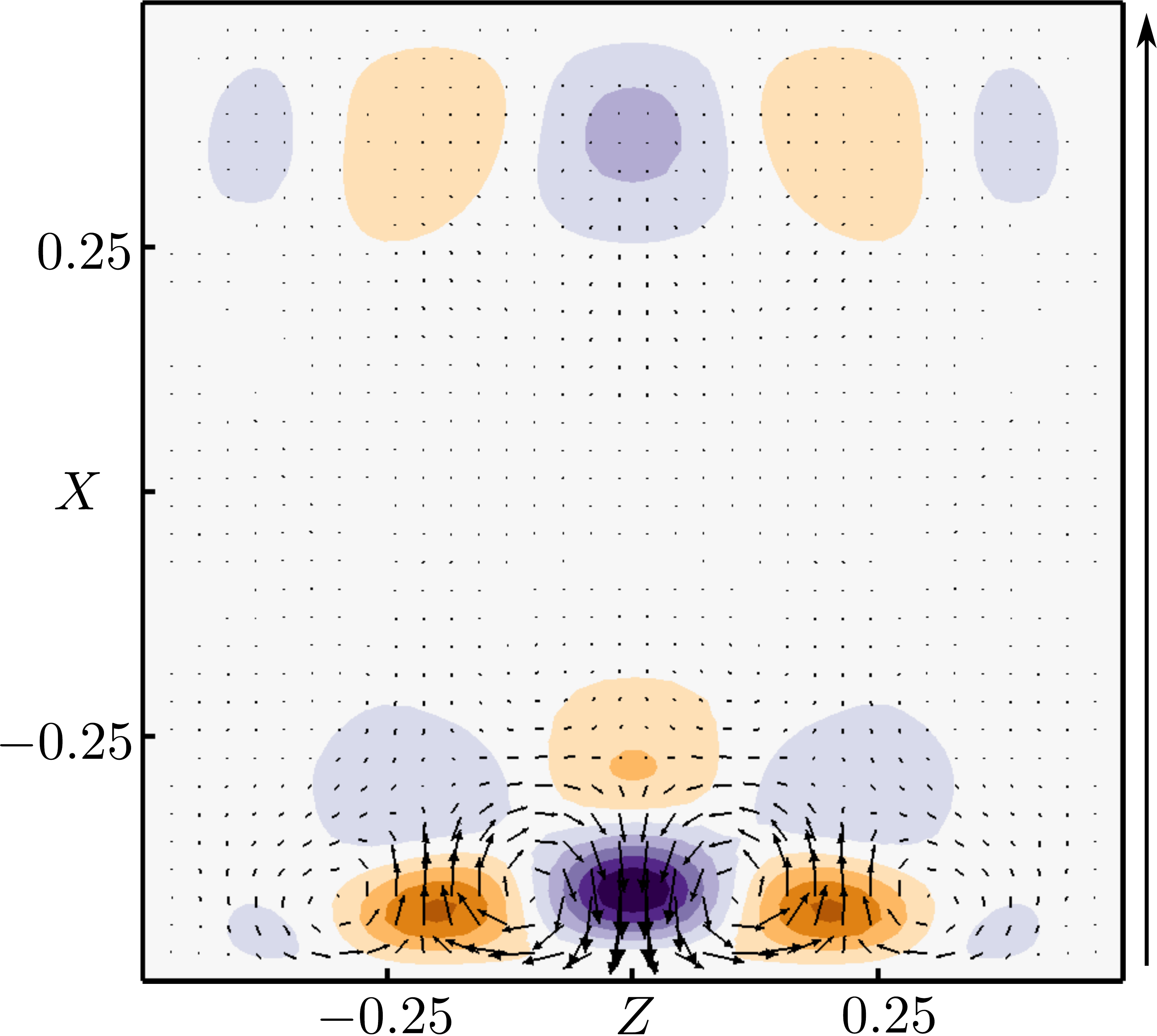}}
  \caption{ (a) Eigenspectrum of the linearized Navier-Stokes operator for the cubical lid-driven cavity flow at $Re = 1930$. (b) Vertical velocity of the unstable eigemode. The isosurfaces depicts ${\bf v} = \pm 10 \%$ of the maximum vertical velocity. (c) Motion it induces in the $y=-0.25$ horizontal plane. Shaded contours highlight the cross-plane velocity while vectors depict the in-plane motion. In both cases, the arrow shows the motion of the lid.}
  \label{fig: Eigenpair}
\end{figure}

Figure~\ref{fig: Eigenpair}(a) shows the eigenspectrum of the linearized Navier-Stokes operator for the cubical lid-driven cavity flow at $Re = 1930$. Only one complex conjugate pair of eigenvalues lies in the upper-half unstable complex plane, characteristic of an Andronov-Poincar\'e-Hopf bifurcation. Comparison of the critical Reynolds number and frequency of the leading eigenvalue ($Re_c = 1914$, $\omega_1 = 0.585$) are in excellent agreement with the values reported by \cite{Feldman_PoF_2010} ($Re_c = 1914$, $\omega=0.575$) and \cite{Kuhlmann_PoF_2014} ($Re_c = 1919.5$, $\omega=0.586$) using non-linear direct numerical simulations. \cite{Kuhlmann_PoF_2014} have furthermore provided some evidences that this Hopf bifurcation is sub-critical.

\bigskip

Figure~\ref{fig: Eigenpair}(b) depicts the vertical velocity component of the leading unstable mode while the motion it induces in the $y=-0.25$ horizontal plane is shown on figure~\ref{fig: Eigenpair}(c). This mode exhibits a mirror symmetry and is made of two structures: counter-rotating vortices along the upstream wall and high- and low-speed streaks along each vertical wall. Such structures, known as Taylor-G\"ortler-like vortices, appear to be reminiscent of the modes in the 2D-periodic lid-driven cavity~\citep{Albensoeder_PoF_2001,Theofilis_JFM_2004,Jerem_CRAS}. This is also suggested by comparing the dominant spanwise wavenumbers $\beta_z$ of the present three-dimensional mode with the characteristics of the $S1$ and $T1$ instability modes from 2D-periodic stability analyses (not shown).

\subsubsection*{Reynolds-Orr analysis}

In order to get a better understanding of how the perturbation extracts its energy from the base flow, the perturbation is decomposed as ${\bf u} = {\bf u}_{\perp} + {\bf u}_{\parallel}$, {\it i.e.} components perpendicular and parallel to the direction of the base flow ${\bf U}_b$~\citep{Albensoeder_PoF_2001}. It highlights that on the one hand the counter-rotating vortices are associated to a motion perpendicular to the base flow's direction (${\bf u}_{\perp}$), while on the other hand the high- and low-speed streaky structure attached to ${\bf u}_{\parallel}$ is everywhere parallel to the direction of the flow. The vortical structure associated to ${\bf u}_{\perp}$ contains roughly 40\% of the perturbation's kinetic energy while the parallel one given by ${\bf u}_{\parallel}$ contains the remaining 60\%. The different physical mechanisms by which the perturbation can extract its energy from the underlying base flow can be unravelled by a careful inspection of the Reynolds-Orr equation. This equation reads
~
\begin{equation}
  \frac{\partial E}{\partial t} = \underbrace{-\int_{\mathrm{V}} {\bf u} \cdot ({\bf u} \cdot \nabla) {\bf U}_b \ \mathrm{dV}}_{\mathrm{Production} \ P} \overbrace{- \frac{1}{Re} \int_{\mathrm{V}} \nabla {\bf u} : \nabla {\bf u} \ \mathrm{dV}}^{\mathrm{Dissipation} \ D}
  \label{eq: Reynolds-Orr}
\end{equation}

\noindent where the first term on the right-hand side is the total production $P$, and the second one is the total dissipation $D$. Introducing the decomposition of the perturbation into this equation, the production term $P$ can be splitted into four different contributions
~
\begin{eqnarray}
    I_1 = - \int_{\mathrm{V}} {\bf u}_{\perp} \cdot ( {\bf u}_{\perp} \cdot \nabla) {\bf U}_b \ \mathrm{dV}, \qquad  I_2 = -\int_{\mathrm{V}}{\bf u}_{\parallel} \cdot ( {\bf u}_{\perp} \cdot \nabla) {\bf U}_b    \  \mathrm{dV}\\
    I_3 = -\int_{\mathrm{V}}{\bf u}_{\perp} \cdot ( {\bf u}_{\parallel} \cdot \nabla) {\bf U}_b \ \mathrm{dV}, \qquad I_4 = -\int_{\mathrm{V}}{\bf u}_{\parallel} \cdot ( {\bf u}_{\parallel} \cdot \nabla) {\bf U}_b   \  \mathrm{dV}.
  \label{eq: production terms}
\end{eqnarray}

\noindent A different physical mechanism is associated to each of these four contributions: $I_2$ is related to the lift-up effect, $I_3$ to the anti lift-up effect, while $I_1$ and $I_4$ are self-induction mechanisms of the vortical and parallel structures, respectively. The sign of the different integrals $I_i$ then informs whether the associated physical mechanism acts as promoting (positive) or quenching (negative) the instability considered. Table~\ref{tab: Reynolds-Orr budget} provides the contribution of each of the different production terms and of dissipation to the total kinetic energy budget of the instability. As can be seen from this table, the rate of extraction of the perturbation's energy from the base flow essentially results from a competition between the destabilizing effect of the lift-up mechanism ($I_2$) and the stabilizing one of dissipation ($D$). The domination of $I_2$ in the energy budget has been related by \cite{Albensoeder_PoF_2001} to the centrifugal nature of the instability considered.

\begin{table}
  \centering
   \begin{tabular}{lllll}
           $I_1$     & $I_2$          & $I_3$     & $I_4$    & $D$ \\
           \hline
           $-0.0146$ & ${\bf 0.8170}$ & $0.1395$ & $0.1033$ & ${\bf -1.0420}$
    \end{tabular}

    \caption{Contribution of the different terms to the total kinetic energy budget of the unstable eigenmode of the cubical lid-driven cavity flow at $Re = 1930$. The sum of the different contributions is equal to $2\sigma = 0.0032$.}
  \label{tab: Reynolds-Orr budget}
\end{table}

\subsection{Non-linear evolution}

\begin{figure}
  \centering
  \includegraphics[width=.75\columnwidth]{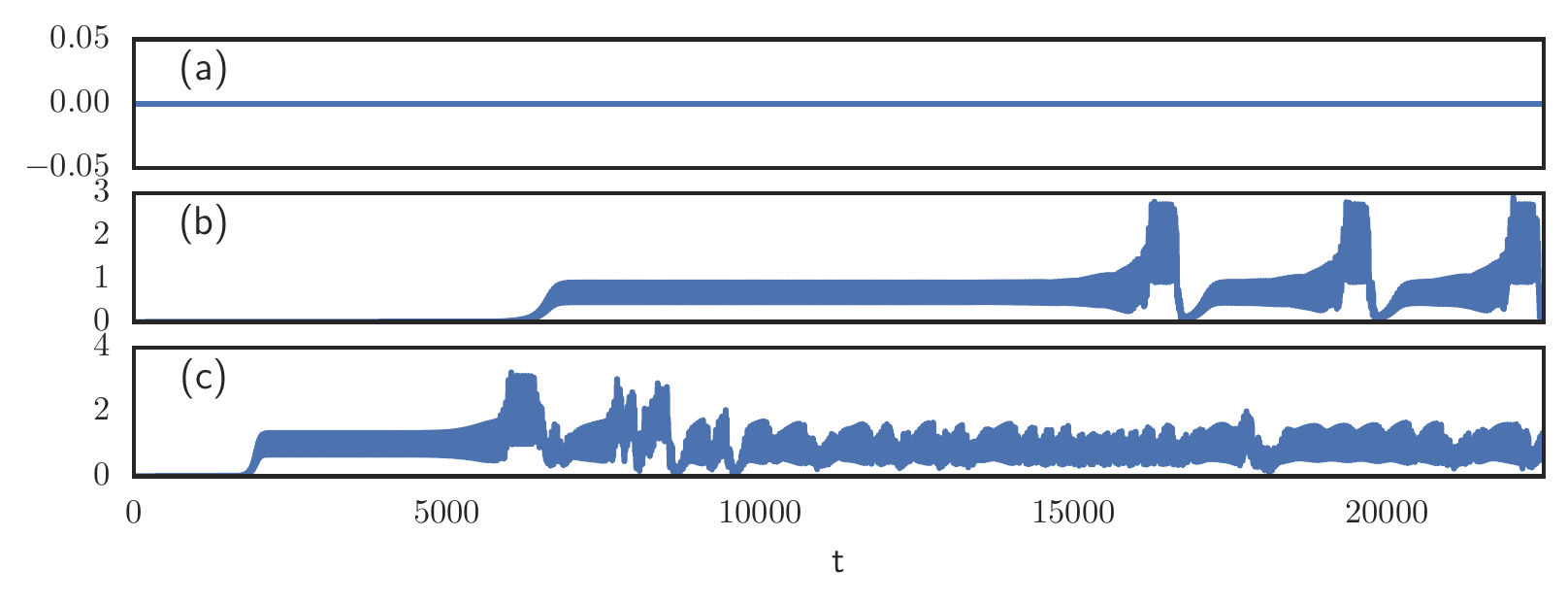}
  \caption{Time-evolution of the perturbation's kinetic energy when increasing the Reynolds number: (a) $Re = 1900$, (b) $Re = 1930$, (c) $Re = 1970$. The energy scale has been multiplied by a factor $10\ 000$.}
  \label{fig: Kinetic energy evolution}
\end{figure}

The time-evolution of the kinetic energy of the perturbation ${\bf u}(t) = {\bf U}(t) - {\bf U}_b$ is depicted on figure~\ref{fig: Kinetic energy evolution} for (a) $Re=1900$, (b) $Re=1930$ and (c) $Re=1970$ respectively. In each case, the direct numerical simulation has been initialized with the appropriate base flow. Figures~\ref{fig: Kinetic energy evolution}(a) and \ref{fig: Kinetic energy evolution}(b) clearly highlight that the fixed point changes from being stable to unstable (in at least along one direction of the phase space) for $1900 < Re_c < 1930$. Once the base flow is linearly unstable, the perturbation escapes exponentially from its vicinity until non-linear saturation yields the perturbation to reach a primary limit cycle (LC1). After a finite but random time, a ''burst'' occurs causing the perturbation to depart away from LC1. It then settles in the vicinity of a different limit cycle for some time before escaping away from it. Such behaviour is known as a chaotic intermittency. What happens after this first intermittency depends on the Reynolds number of the flow as illustrated on figures~\ref{fig: Kinetic energy evolution}(b) and (c). For $Re=1930$ (figure~\ref{fig: Kinetic energy evolution}b), the perturbation comes back in the vicinity of the fixed point before escaping away from it once more and settling on the limit cycle LC1 until a new intermittent event randomly occurs. The process just described then takes place again and again with the time $\tau$ between two intermittencies being apparently random. For $Re=1970$ (figure~\ref{fig: Kinetic energy evolution}c), the dynamics appear as a more chaotic version of those at $Re=1930$. In the rest of this work, attention will essentially be focused on $Re=1930$.

\bigskip

\subsubsection{Temporal analysis}\label{subsubsec: temporal analysis}

\begin{figure}
  \centering
  \includegraphics[width=.75\columnwidth]{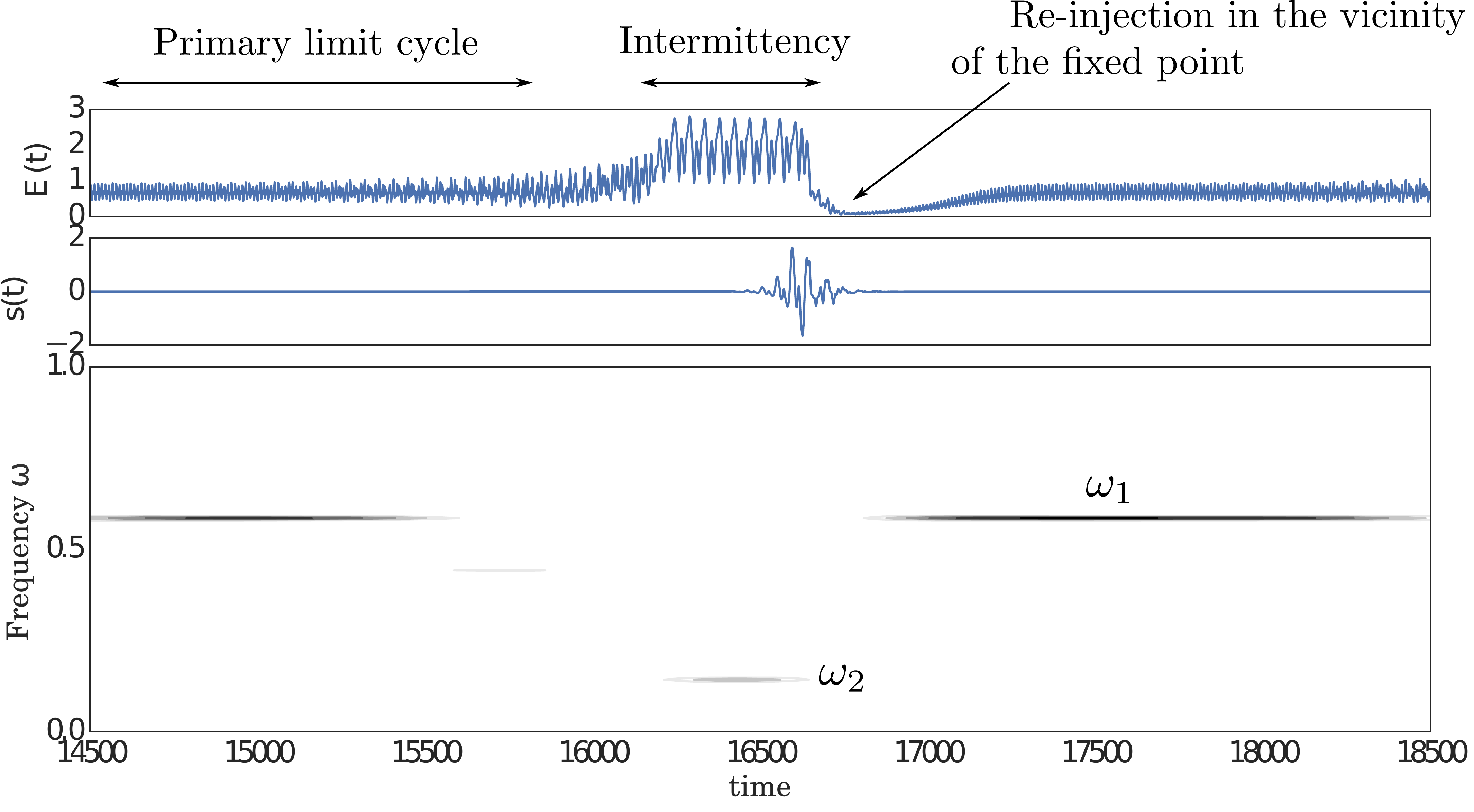}
  \caption{Focus on one intermittent event occuring in the cubical lid-driven cavity flow at $Re = 1930$. The symmetry indicator and energy scales have been multiplied by a factor $100$ and $10\ 000$, respectively.}
  \label{fig: Focus on Re=1930}
\end{figure}

 Figure~\ref{fig: Focus on Re=1930}(a) provides a close-up of the time-evolution of the kinetic energy of the perturbation at $Re=1930$ in the time range $14500 \le t < 18500$. The evolution of the symmetry indicator $s(t)$ is also reported on figure~\ref{fig: Focus on Re=1930}(b). This symmetry indicator is defined as $s(t) = v_1(t) - v_2(t)$, $v_1(t)$ and $v_2(t)$ being the vertical velocity recorded by two probes located at $(x,y,z) = (-0.375,-0.25,\pm 0.25)$, {\it i.e.} symmetrically located with respect to the mid-plane of the cavity. Finally, figure~\ref{fig: Focus on Re=1930}(c) depicts the spectrogram analysis corresponding to the signal from figure~\ref{fig: Focus on Re=1930}(a). Because this signal is non-stationnary, has a time-varying mean value and results from a fundamentally non-linear process, regular Fourier transform or spectrogram analysis out of the box yield unconclusive results. In order to tackle these issues, the {\it Empirical Mode Decomposition}~\citep{Huang_1998} has been used as a pre-processing step in order to obtain a finite and small number of well-behaved components. The python implementation of the MATLAB Time-Frequency Toolbox has then been used to perform the spectrogram analysis. From the different pieces of information scattered between these figures, it appears that

\medskip

\begin{enumerate}

  \item The frequency of the primary limit cycle LC1 is well predicted by the unstable eigenvalue $\sigma_1 \pm i\omega_1 = 1.642 \cdot 10^{-3} \pm i0.585$. Moreover, based on the symmetry measurement given by $s(t)$, LC1 is associated to a mirror-symmetric velocity field.

    \item The frequency observed in the spectrogram during the intermittency seems to be closely related to the eigenvalue $\sigma_2 \pm i\omega_2 = -1.987 \cdot 10^{-2} \pm i0.148$, despite the latter being stable. Moreover, it can be seen from the evolution of $s(t)$ that, during most of the intermittency, the flow exhibits a mirror symmetry as well.

    \item During the process of re-injection of the perturbation in the vicinity of the fixed point, the time evolution of $s(t)$ clearly indicates that the velocity field exhibits a small transient asymmetry. The time scale over which this re-injection occurs is however too small to allow us to characterize any frequency.

\end{enumerate}

\subsubsection{Phase space representation}

\begin{figure}
  \centering
  \subfigure[]{\includegraphics[width=.375\columnwidth]{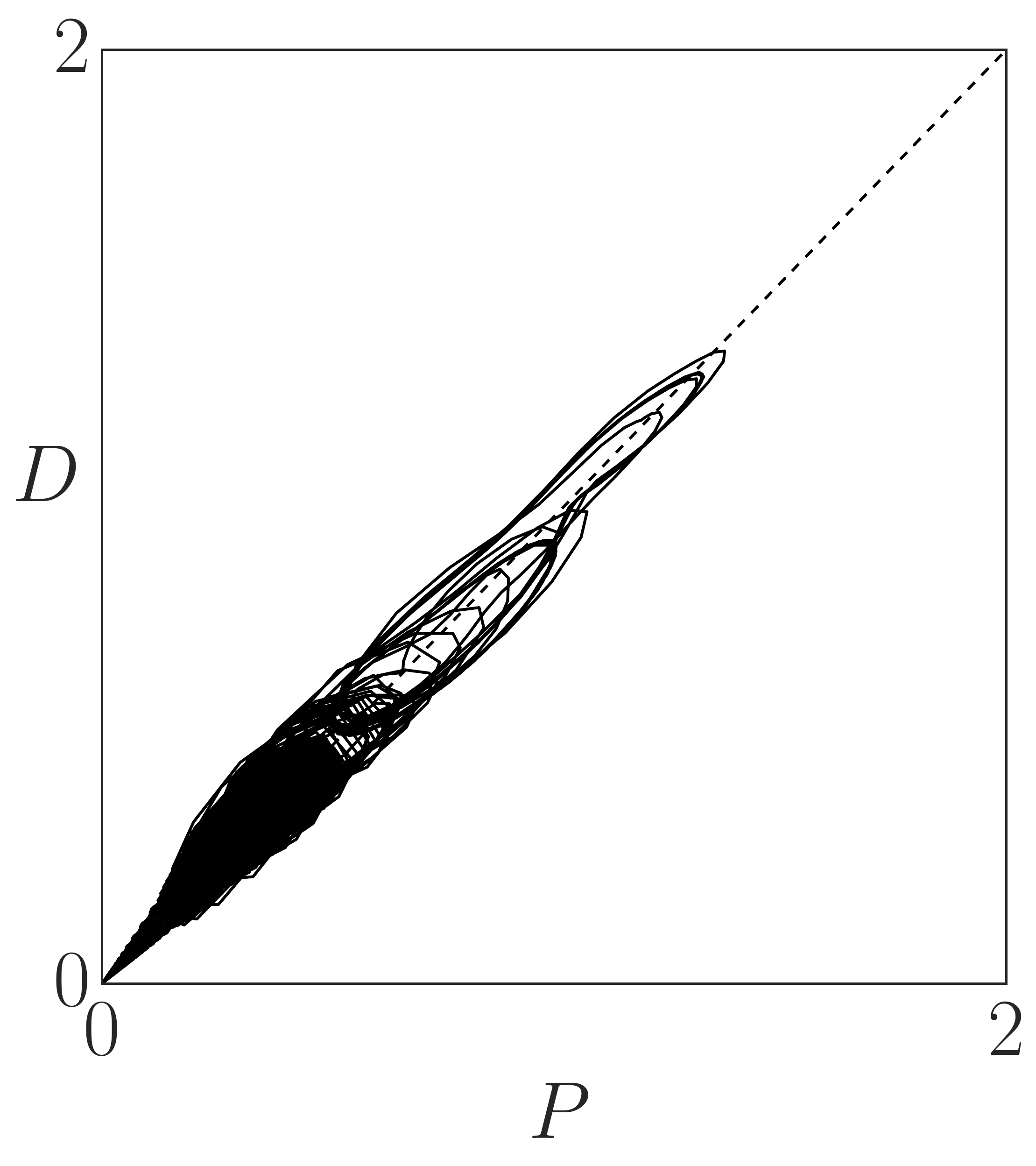}}
  \subfigure[]{\includegraphics[width=.375\columnwidth]{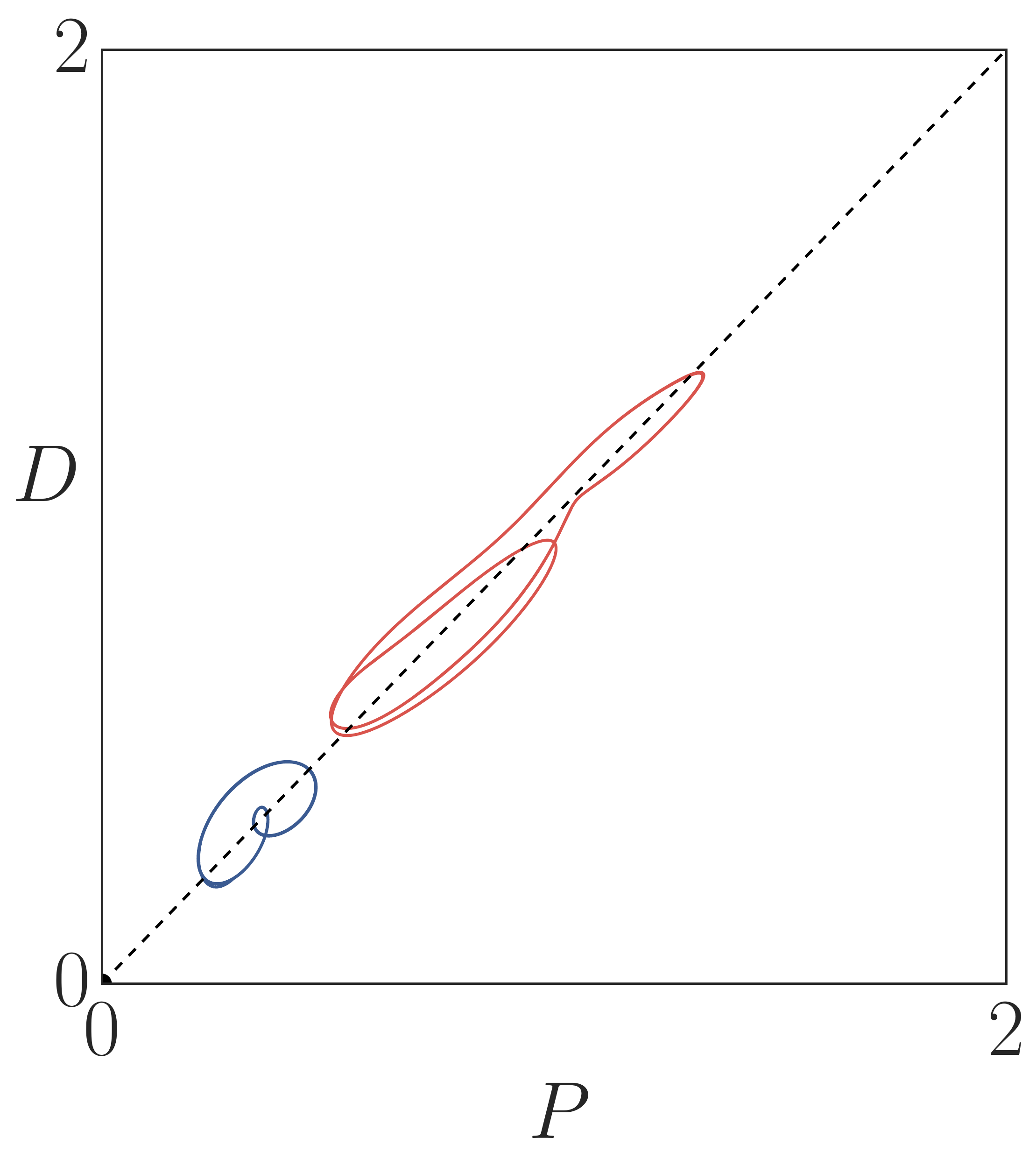}}
  \caption{Projection of the dynamics at $Re = 1930$ in a Production ($P$) vs. Dissipation ($D$) phase diagram. (a) Chaotic dynamics, (b) Dynamics along the two exact coherent states.}
  \label{fig: phase space representation}
\end{figure}

Figure~\ref{fig: phase space representation}(a) depicts the complete dynamics of the system in a production-dissipation diagram. The traces of the intermittent events are clearly visible. Based on the symmetry considerations illustrated on figure~\ref{fig: Focus on Re=1930}(b), a symmetry boundary condition has been applied on the mid-plane of the cavity in order to stabilize the dynamics during the intermittency. Such approach allows us to identify three different {\it exact} solutions :

\begin{itemize}
  \item the original fixed point of the system, located at $(0,0)$ on such diagram,
  \item a low production-dissipation periodic cycle (represented by the blue loop) whose characteristics are well predicted by linear stability analysis,
  \item and a comparatively high production-dissipation periodic cycle (represented by the red loop).
\end{itemize}

\noindent As shown in sections~\ref{subsec: linear stability analysis} and \ref{subsubsec: temporal analysis} using linear stability analysis and DNS, the original fixed point of the system and the limit cycle LC1 are unstable within the mirror-symmetric subspace. Furthemore, DNS with and without symmetry boundary condition have highlighted that the limit cycle LC2 is stable within the mirror-symmetric subspace and unstable within the antisymmetric one. Based on these results, the dynamics of the system observed on figure~\ref{fig: Focus on Re=1930} can be interpreted in the phase space as a trajectory wandering around these three peculiar solutions :

\begin{itemize}
  \item During most of time, the system orbits around the primary limit cycle LC1 (low production-dissipation).
  \item This cycle is however unstable and the system is attracted toward the secondary limit cycle LC2 (high production-dissipation).
  \item This secondary limit cycle is however unstable as well and a re-injection in the vicinity of the original fixed point occurs.
  \item The flow is then slowly driven back to the primary limit cycle due to the linearly unstable nature of the fixed point and the whole process eventually repeats itself.
\end{itemize}

\noindent At the moment, only the linear stability of the fixed point has been investigated thoroughly. The two other key points of this transition scenario are the instability of the two periodic limit cycles. Investigating these properties however requires the use of Floquet stability analysis which is beyond the scope of our present capabilities. Due to the close relationship between the dynamics along the two limit cycles and the characteristics of the leading eigenmodes of the linearized Navier-Stokes operator, it is however believed that the use of a reduced-order model resulting from the projection of the non-linear system onto these leading eigenmodes might help to reduce the complexity of the problem without loss of generality. This reduced-order modeling strategy is currently under development.

\section{Conclusion}
\label{sec: Conclusion}

The linear instability and subsequent transition to chaos of the cubical LDC flow has been investigated by means of global stability analyses and direct numerical simulations. The flow experiences a Hopf bifurcation at a critical Reynolds number $Re_c=1914$ due to a centrifugal instability of the primary vortex core. Analysis of the different production terms highlights that the centrifugal instability can be re-interpreted as a closed-loop instability relying essentially on the lift-up and anti-lift-up mechanisms. Once unstable, the flow is then driven from the unstable steady equilibrium toward a periodic limit cycle. Direct numerical simulations have revealed that, though this periodic limit cycle appears to remain stable over quite a long period of time, intermittent events occuring apparently randomly are eventually observed. The physical origin of these intermittent events is still unknown at the present time. A time-frequency analysis has however shown that the dynamics during the intermittent events appear to be related to one of the least stable eigenvalues of the linearized Navier-Stokes operator. Finally, by imposition of appropriate spatio-temporal symmetries, two periodic limit cycles have been computed. The first one corresponds to the primary limit cycle resulting from the Hopf bifurcation, while the second one appears to govern the dynamics during the intermittent event. Current work aims at getting a better understanding of the transition process between these two states in order to shed some more light on these intermittent chaotic dynamics and the underlying physics.

\subsection*{Acknowledgements}

This work has received the financial support of the French National Agency for Research (ANR) through the grant ANR-09-SYSCOMM-011-SICOGIF. All figures have been produced using the open-source visualization software VisIt~\citep{VisIt} and the python libraries Matplotlib~\citep{matplotlib} and Seaborn.

\bibliography{Main}

\end{document}